\documentclass[
preprint,
 amsmath,amssymb,
 aps,
prc,
]{revtex4-1}
\usepackage{graphicx}
\usepackage{braket}
\usepackage{longtable}

\usepackage{color}

\begin{document}
\title{Low-Energy Gamow-Teller Transitions in deformed $N=Z$ odd-odd Nuclei}
\author{Hiroyuki Morita and Yoshiko Kanada-En'yo}
\affiliation{Department of Physics, Kyoto University, Kyoto 606-8502, Japan.}
\date{\today}

\begin{abstract}
We have investigated the Gamow-Teller transitions from ${}^{22}\textrm{Ne}$ to ${}^{22}\textrm{Na}$
by applying the isospin projected antisymmetrized molecular dynamics combined with generator coordinate method.
We have found that the GT strength from $^AZ(J^\pi T)={}^{22}\textrm{Ne}(0_1^+1)$ is fragmented into two final states ${}^{22}\textrm{Na}(1_{1,2}^+0)$, 
which belong to $K=0$ and $K=1$ bands constructed by a prolately deformed $^{20}$Ne core with a $S=1$ proton-neutron ($pn$) pair. 
Coupling of the intrinsic-spin of the $pn$ pair with the core deformation
play an important role in the GT fragmentation.
The symmetry breaking in the intrinsic-spin rotation 
leads to the $SU(4)$-symmetry breaking
of the $NN$ pair and causes the GT fragmentation.
We have compared the features of the GT transitions 
with those for ${}^{10}\textrm{Be}\to{}^{10}\textrm{B}$ and discuss the 
link of  the $SU(4)$-symmetry  and the GT fragmentation in the deformed systems.
\end{abstract}

\maketitle

\section{Introduction}
Proton-neutron ($pn$) correlation plays important roles in structure properties of $N=Z$ odd-odd nuclei 
(see Ref.~\cite{PNReview} and references therein). 
Gamow-Teller (GT) transition is a good probe for isoscalar $pn$ correlations in $N=Z$ odd-odd nuclei.
Effects of the $pn$ pairing on $B(\textrm{GT})$ have been studied for $N=Z$ odd-odd nuclei in 
light- and medium-mass regions
with a three-body model of two nucleons around spherical cores \cite{GT2}.
One of the interesting features of the GT transitions obtained in $N=Z$ odd-odd nuclei with $LS$-closed core(${}^{4} \textrm{He}$, ${}^{16} \textrm{O}$, ${}^{40} \textrm{Ca}$)
is the remarkably strong GT transition from $J^\pi T=0_1^+1$ to $1_1^+0$ almost exhausting the sum-rule value.
The concentration of the GT transition called the low-energy super GT transitions (LESGT)
is described by the transition between dineutron ($nn$) and deuteron-like $pn$ pairs around the $LS$-closed core 
\cite{LESGT}.

The GT strength functions of $0_1^+1 \to 1_n^+0$ transitions in the $pf$-shell region
have been systematically measured using (${}^{3} \textrm{He}$,$t$) charge-exchange reactions\cite{GT3}.
They have shown that $B(\textrm{GT};0_1^+1 \to 1_1^+0)$ is concentrated on the single $1^+0$ state in the ${}^{42} \textrm{Ca}$ target, 
but, as the mass number increases from $A=42$ to $A=54$, 
the fragmentation of the GT strengths into many $1^+0$ states occurs.
The GT fragmentation in this mass region can be understood by configuration mixing of the particle-hole excited states in the spherical $pf$-shell.

For the GT transitions of ${}^{22} \textrm{Mg}\to{}^{22} \textrm{Na}$, 
the observed strengths show a transient situation between the concentration and fragmentation of the GT transitions. 
Namely, the strength from ${}^{22} \textrm{Mg}(0_1^+1)$ is split into two final states, ${}^{22} \textrm{Na}(1_{1,2}^+0)$, 
in the low-energy region  \cite{DataSheet22}. 
It may imply that a $pn$ pair is formed in ${}^{22} \textrm{Na}(1_{1,2}^+0)$ but it is not the ideal deuteron-like $pn$ pair. 
For the mirror nucleus, ${}^{22} \textrm{Ne}$, a deformed core of the ${}^{16} \textrm{O}+\alpha$ cluster 
with two valence neutrons has been suggested by a theoretical study with the antisymmetrized molecular dynamics (AMD) 
\cite{Ne20AMD,Ne22Cluster,Ne22AMD}. 
${}^{22} \textrm{Mg}$ may also have prolately deformed nature with the spin-isospin 
saturated ${}^{16} \textrm{O}+\alpha$ cluster,  and therefore, the GT transitions of 
${}^{22} \textrm{Mg}\to{}^{22} \textrm{Na}$ should be dominantly 
contributed by the GT transitions of two nucleons at the surface of the deformed nuclei.

The $pn$ pairing in deformed nuclei has been studied with mean-field approaches such as
the generalized Hartree-Fock-Bogoliubov theories(Ref.~\cite{PNDeformation1,PNDeformation2}).
It was pointed out that, in the medium-mass nuclei, 
the pairing correlations reduce because of the nuclear deformation.
Recently, the GT strengths of  ${}^{24} \textrm{Mg}(0_1^+1)\to{}^{24} \textrm{Al}(1^+0)$ have been studied by using the deformed quasiparticle random phase approximation (DQRPA) including $pn$ pairing effects.~\cite{GT1}
They have shown that the GT strengths are scattered in a broad energy region toward the high energy region by introducing the deformation.
Such competitions between deformation and $pn$ pairing should be investigated also in $N=Z$ odd-odd nuclei.

In this paper, we investigate the low-energy GT transitions of 
${}^{22}\textrm{Ne}\to{}^{22} \textrm{Na}$ with the method of the $T$-projected
antisymmetrized molecular dynamics with constraint on the deformation $\beta$ and $\gamma$ parameters 
($T\beta\gamma$-AMD)\cite{Self1}
that can deal with $pn$ correlations in $N=Z$ odd-odd nuclei and quadrupole deformations in light nuclei.
A particular attention is paid on roles of the $pn$ correlation 
in deformation effects on the GT transitions.  We compare the GT transitions
of ${}^{22}\textrm{Ne}\to{}^{22} \textrm{Na}$ with those of ${}^{10} \textrm{Be}\to{}^{10} \textrm{B}$ and discuss
the $pn$ correlation in deformed nuclei and its effect on the GT transitions.

The paper is organized as follows. 
The framework of the present calculation is explained in Sec.~\ref{method}, and 
the calculated results of nuclear properties of energy spectra, $B(M1)$, $B(E2)$, and $B(\textrm{GT})$, are shown in Sec.~\ref{results}. 
single-particle properties and dependence of $B(\textrm{GT})$ on deformation are discussed in Sec.~\ref{discussion}. 
A summary and an outlook are given in Sec. \ref{summary}.

\section{Method}
\label{method}
\subsection{$T\beta\gamma$-AMD}
We applied the $T\beta\gamma$-AMD method to calculate ${}^{22} \textrm{Na}$ and 
${}^{10} \textrm{B}$. The method was constructed in order to study $pn$ correlations in  $N=Z$ odd-odd nuclei,
and applied for a study of GT transitions of ${}^{10} \textrm{Be}\to{}^{10} \textrm{B}$. 
It is an useful approach to describe $pn$ correlations in deformed nuclei because 
the method can control the isospin ($T=0,1$) and the collective deformation $\beta$ and $\gamma$ of $N=Z$ odd-odd nuclei, simultaneously.
In this section, we briefly explain the mathematical formulations of the  $T\beta\gamma$-AMD.
For detailed formulations, the reader is referred to Ref.~\cite{Self1}.


The $T\beta\gamma$-AMD is based on the antisymmetrized molecular dynamics (AMD), 
in which Slater determinants of Gaussian wave packets are used as basis wave functions:
\begin{equation}
\Ket{\Phi\left(\beta,\gamma\right)} = \mathcal{A} \left[\Ket{\phi_1} \Ket{\phi_2} \cdots \Ket{\phi_A}\right],
\end{equation}
\begin{equation}
\Ket{\phi_i} = \left(\frac{2\nu}{\pi}\right)^{\frac{3}{4}} \exp\left[-\nu\left({\mbox{\boldmath $r$}}_i-\frac{{\mbox{\boldmath $Z$}}_i}{\sqrt{\nu}}\right)^2\right] \Ket{\mbox{\boldmath $\xi$}_i}\Ket{\tau_i}.
\end{equation}
Here, we use $\nu = 0.16$~($\textrm{fm}^{-2}$) for ${}^{22} \textrm{Na}$, which reproduces radii of $sd$-shell nuclei.
In order to obtain the optimized wave functions for parity $(\pi)$ and isospin ($T$) eigen states, 
the $\pi$- and $T$-projections are performed before energy variation:
\begin{equation}
\Ket{\Phi^{\pi T}\left(\beta,\gamma\right)} = \hat{P}^\pi\hat{P}^T \Ket{\Phi\left(\beta,\gamma\right)},
\end{equation}
where $\hat{P}^\pi$ and $\hat{P}^T$ are parity and isospin projection operators, respectively.
For the ${}^{\pi}T$-projected wave function, we perform the energy variation under the constraints on the deformation parameters, $\beta$ and $\gamma$,
so as to obtain the optimized states corresponding to each ($\beta$,$\gamma$).
After the variation, the obtained wave functions $\Ket{\Phi^{\pi T}\left(\beta,\gamma\right)}$ are 
projected onto the total angular momentum $J$ eigen states; $\hat{P}_{MK}^J \Ket{\Phi^{\pi T}\left(\beta,\gamma\right)}$.
Here $\hat{P}_{MK}^J$ is the angular momentum projection operator.
Furthermore, we superpose these $J^\pi T$ eigenstates over the ($\beta$,$\gamma$) plane with the generator coordinate method (GCM) to take into account the quantum fluctuations for quadrupole deformations:
\begin{equation}
\Ket{J^\pi_n T;M} = \sum_{iK} c^{iK}_n \hat{P}_{MK}^J \Ket{\Phi^{\pi T}\left(\beta_i,\gamma_i\right)}.
\end{equation}
Here, the parameters, $\beta$ and $\gamma$, are treated as generator coordinates in the GCM, and the $K$-mixing is taken into account. 
We call this method the $T\beta\gamma$-AMD+GCM. 

\subsection{Effective interactions}
We use the Hamiltonian
\begin{equation}
H = K-K_\textrm{cm} + V_\textrm{c} + V_{LS} + V_\textrm{Coulomb},
\end{equation}
where $K$ is the kinetic energy, $K_\textrm{cm}$ is the kinetic energy of the center of mass,
and $V_\textrm{c}$, $V_{LS}$, and $V_\textrm{Coulomb}$ are the central force, the spin-orbit force, and the Coulomb force, respectively. 
For the central force, the Volkov No.~2 force \cite{Volkov} with the Majorana exchange parameter $m=0.6$
is used.
The Bartlett and Heisenberg parameters $b=h=0.06$ are adopted, which are phenomenologically adjusted to the energy difference between the lowest $T=0$ and $T=1$ states in ${}^{10} \textrm{B}$ \cite{Self1}. 
For $V_{LS}$, we use the spin-orbit part of the G3RS force \cite{G3RS1,G3RS2}
with the same strength parameters $u_{ls}=u_1=-u_2=1300$~MeV as those used in the previous works \cite{Self1,Self2}.

\section{Results}
\label{results}

\begin{figure}
\includegraphics[width=\hsize]{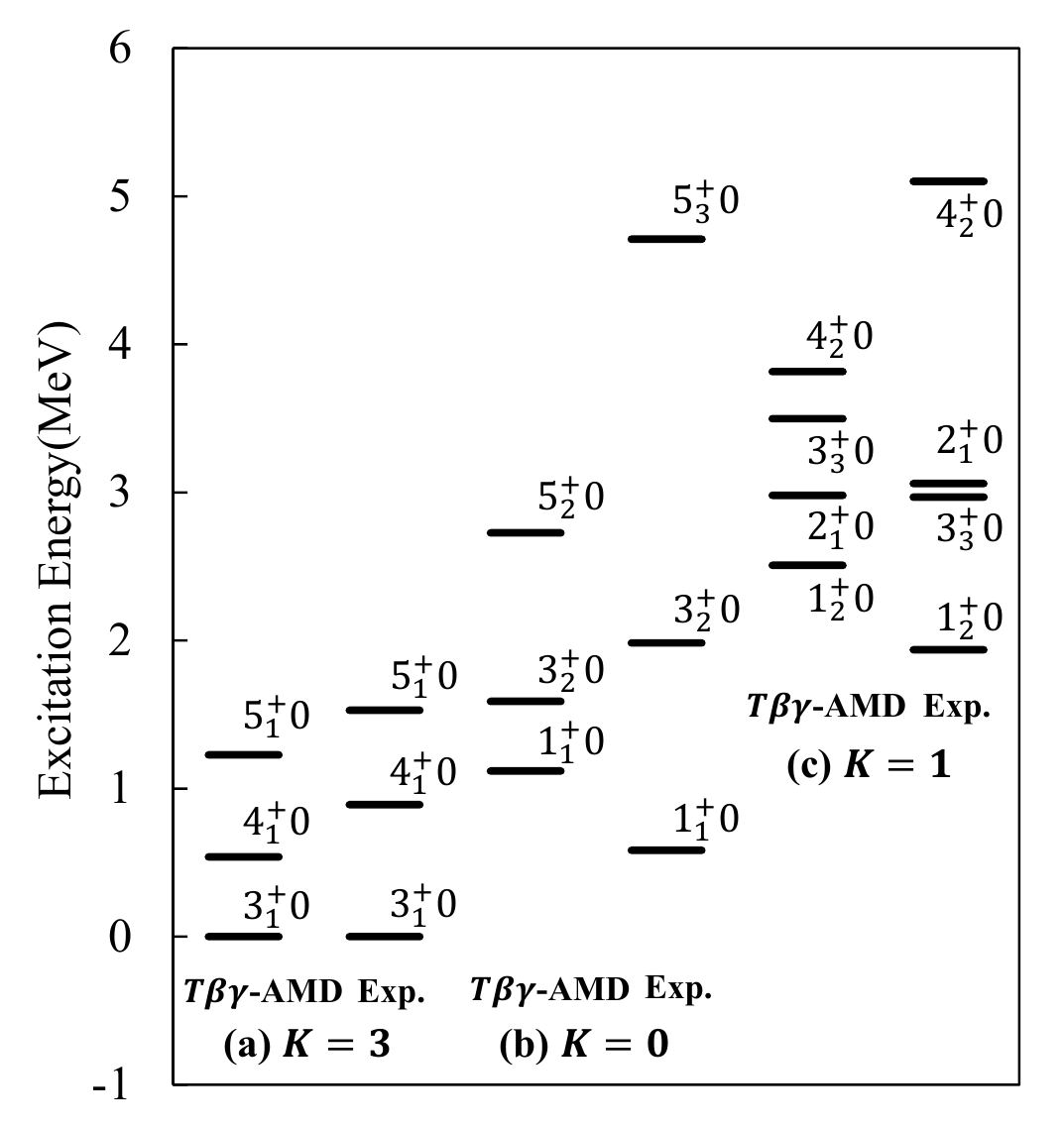}
\caption{The low-lying spectra of ${}^{22} \textrm{Na}$ in the $T=0$ $K=0,1,3$ bands. 
Calculated and experimental spectra are shown in the left and right, respectively. 
The experimental data are taken from \cite{DataSheet22}.}
\label{NaT0Spec}
\end{figure}

\begin{figure}
\includegraphics[width=\hsize]{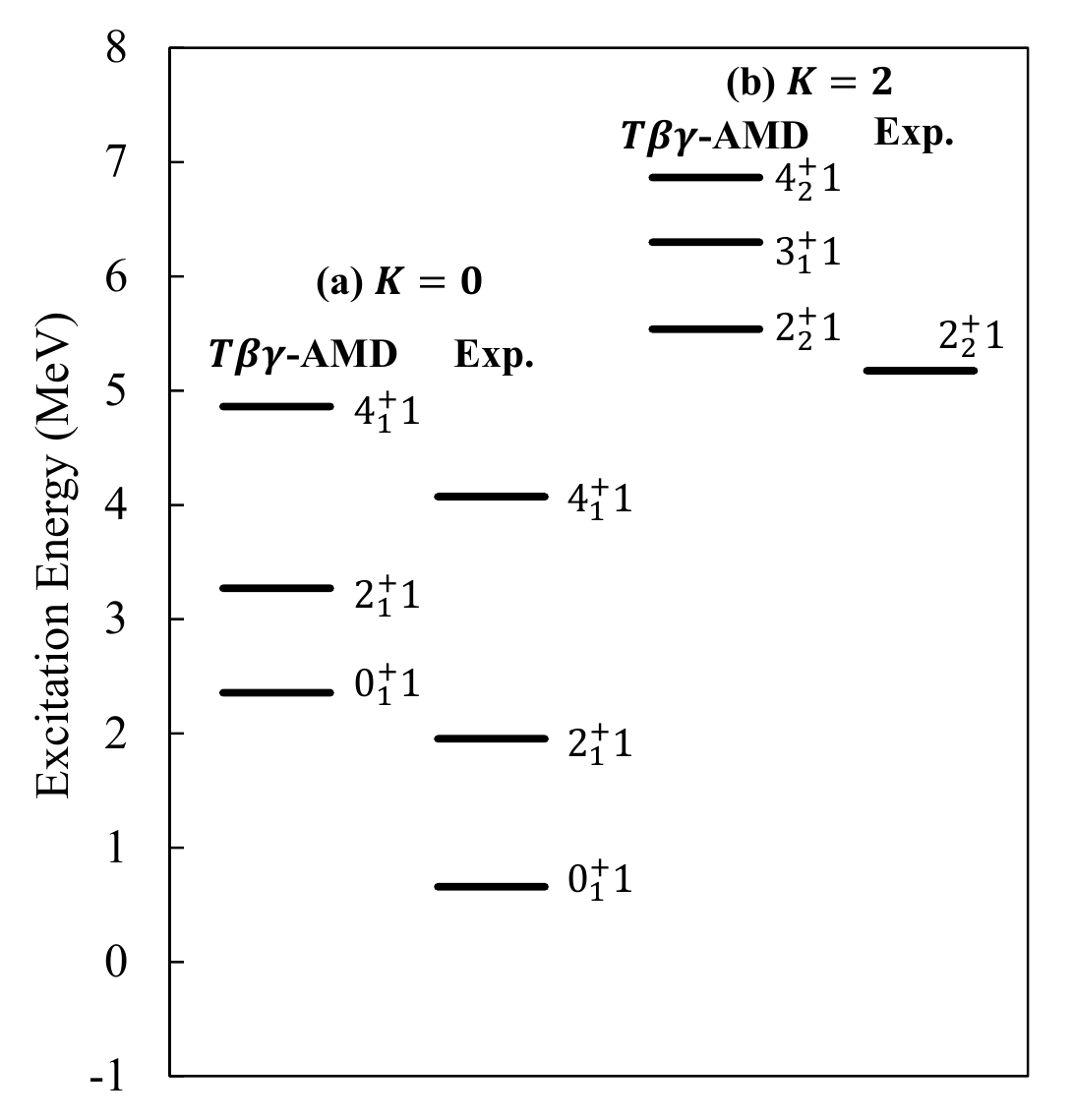}
\caption{The low-lying spectra of ${}^{22} \textrm{Na}$ in the $T=1$ $K=0,2$ bands. For each band,
calculated and experimental spectra are shown in the left and right, respectively. The experimental data are taken from \cite{DataSheet22}.}
\label{NaT1Spec}
\end{figure}

\begin{figure}
\includegraphics[width=\hsize]{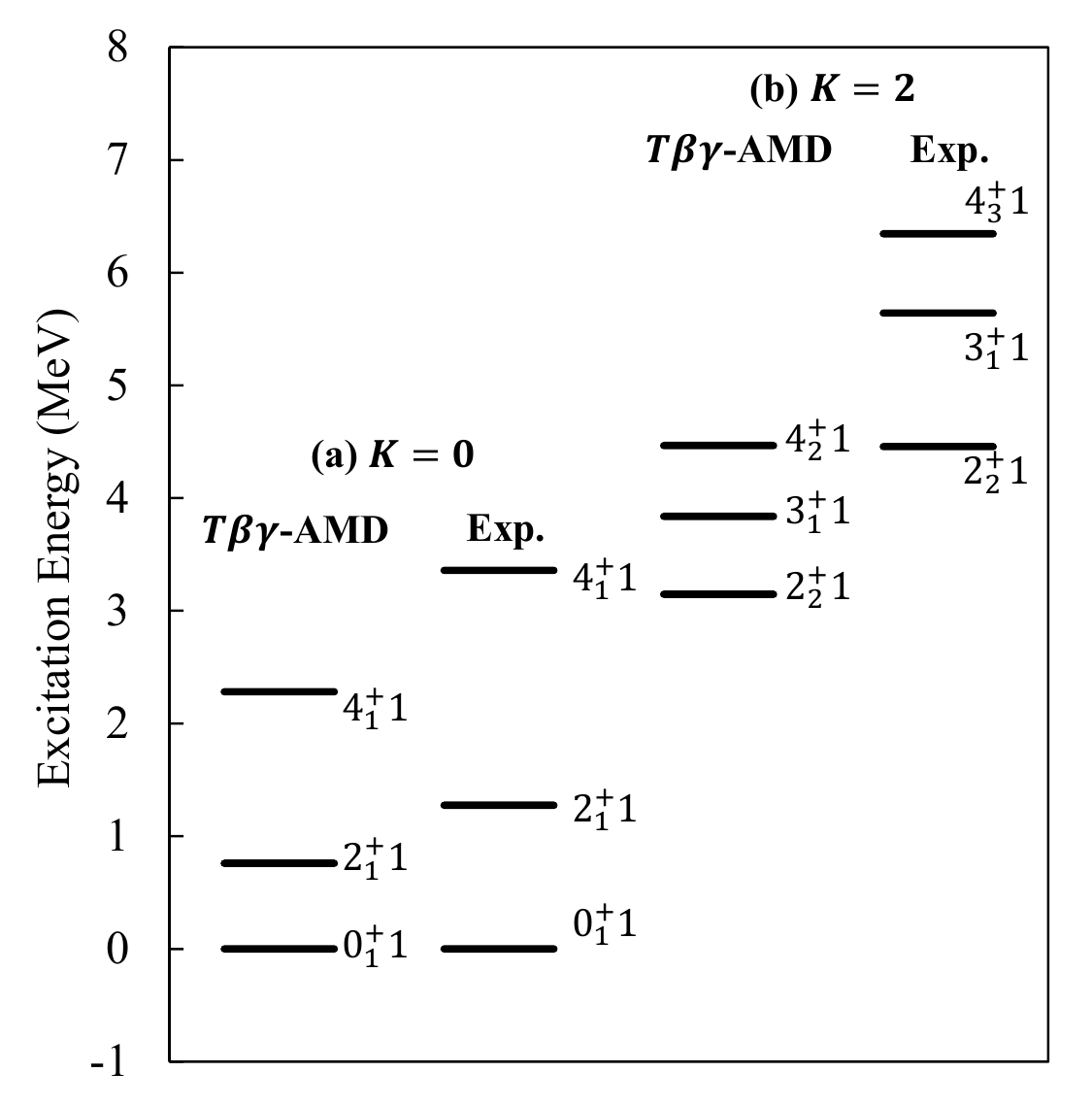}
\caption{The low-lying spectra of ${}^{22} \textrm{Ne}$ in the $K=0,2$ bands. For each band, 
calculated and experimental spectra are shown in the left and right, respectively.
The experimental data are taken from \cite{DataSheet22}.}
\label{NeSpec}
\end{figure}

\subsection{Energy spectra and electro-magnetic transitions in ${}^{22} \textrm{Na}$}
The energy spectra for the $E<8$~MeV states in ${}^{22} \textrm{Na}$ are shown in Fig.~\ref{NaT0Spec} and ~\ref{NaT1Spec} and those in ${}^{22} \textrm{Ne}$ are shown in Fig.~\ref{NeSpec}.
The calculation reasonably reproduces low-lying energy spectra. In particular, the spectra of $J\le 3$ states agree with the
experimental data.



\begin{longtable}{ccc}
\caption{The electric and magnetic moments and transition strengths 
in ${}^{22} \textrm{Na}$. The calculated $Q$ ($e$\,$\textrm{fm}^2$) and $\mu$  ($\mu_N$) moments, 
$B(E2)$ ($e^2$\,$\textrm{fm}^4$), and $B(M1)$ ($\mu_N^2$) are shown together with the experimental data
from \cite{DataSheet22}. The binding energy (MeV) of the ground state ${}^{22} \textrm{Na}(3^+_1 0)$ is 
also shown.
}
\label{EMprop}
\\
\hline\hline
Observable&$T\beta\gamma$-AMD+GCM&Exp\\
\hline
binding energy&173.041&174.1456\\
$Q(3_1^+0)$ &17.66&18.0(11)\\
$\mu(3_1^+0)$&1.784&1.746(3)\\
$\mu(1_1^+0)$&0.622&0.535(10)\\
\hline
$K=3$&\\
\hline
$B(E2;5_1^+0\to 4_1^+0)$&49.9&58(18)\\
$B(E2;4_1^+0\to 3_1^+0)$&56.8&91(3)\\
$B(E2;5_1^+0\to 3_1^+0)$&12.0&19.0(15)\\
$B(M1;4_2^+1\to 5_1^+0)$&2.29&--\\
$B(M1;3_1^+1\to 4_1^+0)$&3.35&--\\
$B(M1;2_2^+1\to 3_1^+0)$&3.97&--\\
\hline
$K=0$&\\
\hline
$B(E2;3_2^+0\to 1_1^+0)$&35.5&69(7)\\
$B(E2;5_2^+0\to 3_2^+0)$&41.2&51(22); $5_3^+0\to 3_2^+0$\\
$B(M1;0_1^+1\to 1_1^+0)$&5.00&4.96(18)\\
$B(M1;2_1^+1\to 3_2^+0)$&3.28&--\\
$B(M1;4_1^+1\to 3_2^+0)$&0.27&$>5.37$\\
$B(M1;4_1^+1\to 5_2^+0)$&3.02&2.2(9); $4_1^+1\to 5_3^+0$\\
\hline
$K=1$&\\
\hline
$B(E2;2_1^+0\to 1_2^+0)$&43.9&--\\
$B(E2;3_3^+0\to 2_1^+0)$&10.2&--\\
$B(E2;3_3^+0\to 1_2^+0)$&13.9&--\\
$B(E2;4_2^+0\to 3_3^+0)$&14.0&--\\
$B(E2;4_2^+0\to 2_1^+0)$&24.4&--\\
$B(M1;0_1^+1\to 1_2^+0)$&4.12&4.3(13)\\
$B(M1;2_1^+1\to 2_1^+0)$&2.21&1.22(16)\\
\hline
$K=0,1$ inter-band &\\
\hline
$B(E2;2_1^+0\to 1_1^+0)$&4.1&0.10(7)\\
$B(E2;3_2^+0\to 1_2^+0)$&1.7&--\\
$B(E2;1_1^+0\to 1_2^+0)$&7.81&--\\
$B(E2;2_1^+0\to 3_2^+0)$&9.42&--\\
\hline\hline
\end{longtable}

The $K=0,1,$ and $3$ bands are obtained in the isoscalar ($T=0$) states of ${}^{22} \textrm{Na}$.
The ground band is the $K=3$ band consisting of the $3_1^+0$, $4_1^+0$ and $5_1^+0$ states.
As shown in Table~\ref{EMprop}, $E2$ transitions, $5_1^+0\to 4_1^+0$, $4_1^+0\to 3_1^+0$, and $5_1^+0\to 3_1^+0$, in the  $K=3$ band
are strong because of the prolate deformation consistently with the experimental data.
The band head $3_1^+0$ obtained with the GCM has the largest overlap as 89.7\% with the 
$J^\pi T=3^+ 0$ and $K=3$ projected basis 
wave function, $\hat{P}_{M3}^{J=3} \Ket{\Phi^{+0}\left(\beta,\gamma\right)}$, at $\beta=0.29$ and $\gamma=0.19$
on the $\beta$-$\gamma$ plane, and shows the $K=3$ nature of the 
deformed band. 
As is discussed later, this band is understood as the deformed ${}^{20} \textrm{Ne}$ core with a spin-aligned pair of 
a proton and a neutron in the lowest valence orbit.

The $1_1^+0$ state is the band head of the $K=0$ band with the rotational band members $1_1^+0$, $3_2^+0$ and $5_2^+0$
with the strong $E2$ transitions of $5_2^+0\to 3_2^+0$ and $3_2^+0\to 1_1^+0$.
The experimental value of $B(E2;3_2^+0\to 1_1^+0)$ is consistent with the calculated value.
For the $5_2^+0$ state, we tentatively assign the experimental $5_3^+0$ state because 
the experimental $B(E2;5_3^+0\to 3_2^+0)$ is the same order of the calculated value of $B(E2;5_2^+0\to 3_2^+0)$.
The band head $1_1^+0$ state has the largest overlap (82.7\%) with the $J^\pi T= 1^+0$ and $K=0$ projected wave function at 
$\beta=0.31$ and $\gamma=0.11$ and shows the $K=0$ nature of the deformed band.
Compared with the features of the ground $K=3$ band, 
it can be seen that the $K=0$ band has the similar deformation 
but has different nature of the spin configuration of the valence proton and neutron.

The $1_2^+0$ and $2_1^+0$ states can be classified in the $K=1$ band members because of the 
strong $E2$ transition and the $K=1$ nature. The band head $1_2^+0$ state has 
the largest overlap (76.6\%) with the $J^\pi T= 1^+0$ and $K=1$ projected 
wave function at $\beta=0.29$ and $\gamma=0.19$ and almost the same intrinsic deformation as 
those of the $K=0$ and $3$ bands.

It is worth to discuss the inter-band $E2$ transitions to establish the $K=0$ and $1$ bands.
The calculated $B(E2)$ values for the inter-band transitions 
$1_2^+0\to 1_1^+0$, $2_1^+0\to 3_2^+0$, $2_1^+0\to 1_1^+0$ and $1_2^+0\to 3_2^+0$ between $K=0$ and $1$ bands
are generally small. The experimental data of $B(E2;2_1^+0\to 1_1^+0)$ are small and support our results.

The $M1$ transition is a good probe for spin configuration because it is contributed by the spin flip transitions.
Similarly to the GT transitions, the isovector $M1$ transitions from $T=1$ states are useful observables 
for spin structure in $T=0$ states. In Table \ref{EMprop}, the calculated strengths of the isovector $M1$ transitions
in ${}^{22} \textrm{Na}$ are shown together with the experimental data.
The observed strong $M1$ transitions from the $T=1$ $K=0$ bands are described well by the present
calculation except for $B(M1;4_1^+1\to 3_2^+0)$. 
The calculated $B(M1;4_1^+1\to 5_2^+0)$ in our calculation is comparable with the experimental $B(M1;4_1^+1\to 5_3^+0)$. It supports
our assignment of the calculated $5_2^+0$ to the experimental $5_3^+0$.
For the transition $4_1^+1\to 3_2^+0$,  the present calculation underestimates the experimental $B(M1)$.
This tendency is also seen in the shell model calculation \cite{M1ShellModel}.

\subsection{Gamow-Teller transitions and $SU(4)$-symmetry}

\begin{table}
\caption{The GT transition strengths defined by Eq.~\eqref{GTdef}. of ${}^{22} \textrm{Ne}\to {}^{22} \textrm{Na}$. The experimental data are taken from \cite{DataSheet22}.}
\label{GTprop}
\begin{ruledtabular}
\begin{tabular}{ccc}
Observable&$T\beta\gamma$-AMD+GCM&Exp\\
\hline
$K=2\to K=3$&&\\
\hline
$B(\textrm{GT};4_2^+1\to 5_1^+0)$&0.95&--\\
$B(\textrm{GT};3_1^+1\to 4_1^+0)$&1.27&--\\
$B(\textrm{GT};2_2^+1\to 3_1^+0)$&1.51&--\\
\hline
$K=0\to K=0$&&\\
\hline
$B(\textrm{GT};0_1^+1\to 1_1^+0)$&1.98&(0.949(28))\\
$B(\textrm{GT};2_1^+1\to 1_1^+0)$&0.30&--\\
$B(\textrm{GT};2_1^+1\to 3_2^+0)$&1.24&--\\
$B(\textrm{GT};4_1^+1\to 5_2^+0)$&1.12&--\\
\hline
$K=0\to K=1$&&\\
\hline
$B(\textrm{GT};0_1^+1\to 1_2^+0)$&1.55&(1.43(8))\\
$B(\textrm{GT};2_1^+1\to 1_2^+0)$&0.37&--\\
$B(\textrm{GT};2_1^+1\to 2_1^+0)$&0.82&--\\
$B(\textrm{GT};4_1^+1\to 3_2^+0)$&0.12&--\\
\end{tabular}
\end{ruledtabular}
\end{table}

The GT transition operator is given as:
\begin{equation}
\label{GTdef}
B\left(\textrm{GT}\right) = \frac{1}{2J_i+1}\left|\Braket{J_f||\sum_i {\mbox{\boldmath $\sigma$}}^i\tau^i_\pm||J_i}\right|^2, 
\end{equation}
which changes the spin and isospin of the initial state as $\Delta S=1$ and $\Delta T=1$, and is regarded as 
the rotation operator in the spin and isospin $SU(4)$-space.
In the previous work \cite{Self2}, we have investigated the GT transitions of  ${}^{10} \textrm{Be}\to {}^{10} \textrm{B}$ and
discussed the spin-isospin partner states connected with strong GT transitions in ${}^{10} \textrm{Be}$ and ${}^{10} \textrm{B}$.
For the assigned partner states, the initial and final states are described by the 
$S=0,T=1$ $nn$ and $S=1,T=0$ $pn$ pairs of valence two nucleons around the $2\alpha$ core, respectively.
It means that the strong GT transitions are understood by the transitions of the $NN$ pairs with the approximate
$SU(4)$-symmetry in the spin and isospin space.
The $SU(4)$-symmetry of the $NN$ pair is partially broken in ${}^{22} \textrm{Na}$, because intrinsic spin of the $NN$ pairs 
strongly couples to the core deformation because of the spin-orbit mean potential and therefore 
the symmetry of the spin rotation is broken. 
Nevertheless, we can also assign the spin-isospin partners in the GT transitions of ${}^{22} \textrm{Ne}\to{}^{22} \textrm{Na}$ 
for the sub spaces of the final states, which are separated by the deformation effect.
Below, we discuss the GT transitions and assignments of spin-isospin partners in ${}^{22} \textrm{Ne}\to{}^{22} \textrm{Na}$.
%

The calculated GT transitions of ${}^{22} \textrm{Ne}\to{}^{22} \textrm{Na}$ are shown in 
Table \ref{GTprop}.
We obtained the significant GT transition strengths from the $K=0$ and $K=2$ bands  of ${}^{22} \textrm{Ne}$ to 
the $K=0$, $K=1$, and $K=3$ bands of ${}^{22} \textrm{Na}$. 

The GT transition strengths from the 
$K=0$ band states of ${}^{22} \textrm{Ne}$ are split into the $K=0$ and $K=1$ band states of ${}^{22} \textrm{Na}$. 
The GT transition strengths from the $0_1^+1$ are fragmented into 
two low-lying $1^+ 0$ states, ${}^{22} \textrm{Na}(1^+_10)$ and ${}^{22} \textrm{Na}(1^+_20)$.
The result is consistent with the experimental observations for the mirror transitions 
${}^{22} \textrm{Mg}(0_1^+1)\to{}^{22} \textrm{Na}(1_{1,2}^+0))$.
Also for the transitions from the initial $2_1^+1$, we obtain the GT strengths fragmented into $2_1^+0$ and $3_2^+0$.
These final states in the  $K=0$ and $K=1$ bands in ${}^{22} \textrm{Na}$ are regarded as spin-isospin partners of the 
initial $K=0$ band states in ${}^{22} \textrm{Ne}$. 
As we show in detail later, the $K$ quanta, $K=0$ and $K=1$, of the final states in ${}^{22} \textrm{Na}$ 
are mainly contributed by the intrinsic spin of the $T=0$ $S=1$ $pn$ pair; $S_z=\pm 1$ contributes to $K=\pm 1$ and $S_z = 0$ corresponds to $K=0$.
The GT transitions into the former and the latter bands occur by the 
spin flip $\sigma_\pm\propto \sigma_x\pm i\sigma_y$ with $\Delta S_z=\pm 1$ and non-flip operators $\sigma_0$
with $\Delta S_z=0$, respectively, because the initial state in the $K=0$ band has the dominant
$S_z=0$ component. 
It means that the splitting of the GT transition strengths is a consequence 
of the formation of two low-lying bands, $K=0$ and $K=1$, because of the 
$T=0$ $S=1$ $pn$ pair correlation in the deformed system.
In other words, because of the symmetry breaking of the spin rotation of the $T=0$ $S=1$ $pn$ 
pair in ${}^{22} \textrm{Na}$, the GT transition strengths from the ${}^{22} \textrm{Ne}(0^+_11)$
does not concentrate to a single $1^+0$ state. It is a different situation from the super-allowed GT transitions of 
$^6\textrm{Li}$ and $^{10}\textrm{B}$, in which 
the spin $S=1$ of the $pn$ pair couples weakly with the core and approximately maintains the  $SU(4)$-symmetry.

For the GT transition to the ground $K=3$ band in ${}^{22} \textrm{Na}$,  
the strong GT transitions from $K=2$ band in  ${}^{22} \textrm{Ne}$ are obtained:
the strengths from the initial states, $2_2^+1$, $3_1^+1$, and $4_2^+1$, concentrate into the final states, $3_1^+0$, $4_1^+0$, $5_1^+0$, respectively.
In the initial states, the quanta $K=2$ are given by the orbital angular momenta of the valence $S=0$ $nn$ pair. 
The $K=3$ of the final states are described by the orbital angular momentum $L_z=2$ and  
the spin $S=1$ of the $pn$ pair aligned to the $z$ direction of the deformed intrinsic state
because of the spin-orbit mean potential.
The GT transitions $K=2\to K=3$ occur as the transition $S=0$ $nn$ $\to$ $T=0$ $S=1$ $pn$ with $\Delta S=1$
by the spin flip operator $\sigma_\pm$.
Therefore, the $K=3$ states in ${}^{22} \textrm{Na}$ 
are assigned to spin-isospin partners of the $K=2$ states in ${}^{22} \textrm{Ne}$.  

\section{Discussion}
\label{discussion}
In this section, we analyze single-particle orbits of valence protons and neutrons in $^{22}\textrm{Na}$
and compare the GT transitions of $^{22}\textrm{Ne}\to ^{22}\textrm{Na}$ with those of $^{10}\textrm{Be}\to ^{10}\textrm{B}$. 

\subsection{Single-particle orbit and Nilsson diagram} 
We discuss single-particle properties of the $K=0,1,3$ bands of ${}^{22} \textrm{Na}$ analyzing the major components of the band head states.
In Table \ref{NaSP}, we show single-particle properties of the intrinsic wave function at $(\beta,\gamma)=(0.29,0.19)$,
which is the dominant component of the ground $3_1^+0$ state for the $K=3$ band. 
The single-particle energies, the expectation values of squared angular momenta
and orbital angular momenta, and positive parity probabilities are shown. In order to discuss the link with Nilsson orbits, 
we also show the $\Omega$ and $\Lambda$ values for each single-particle orbit,  
\begin{equation}
\Omega=\sqrt{\Braket{\phi^\textrm{s.p.}_i|\hat{j}_z^2|\phi^\textrm{s.p.}_i}},
\end{equation}
\begin{equation}
\Lambda=\sqrt{\Braket{\phi^\textrm{s.p.}_i|\hat{\ell}_z^2|\phi^\textrm{s.p.}_i}}.
\end{equation}


\begin{table*}
\caption{
The single-particle properties of the major component of the ${}^{22} \textrm{Na}(3_1^+0)$ ground state at $(\beta,\gamma)=(0.29,0.19)$. 
The column labeled ``parity'' stands for the fraction of positive parity component in each single-particle state.}
\label{NaSP}
\begin{ruledtabular}
\begin{tabular}{ccccccccccccc}
\multicolumn{6}{c}{neutron}&\multicolumn{6}{c}{proton}&\\
\cline{1-6}\cline{7-12}
Energy&$\langle\hat{j}^2\rangle$&$\langle\hat{\ell}^2\rangle$&parity&$\Omega$&$\Lambda$&Energy&$\langle\hat{j}^2\rangle$&$\langle\hat{\ell}^2\rangle$&parity&$\Omega$&$\Lambda$&shell\\
\cline{1-6}\cline{7-12}\cline{13-13}
$-60.94$&0.75&0.00&1.00&0.50&0.02&$-55.98$&0.75&0.00&1.00&0.50&0.02&$s_{1/2}$\\
$-59.29$&0.75&0.00&1.00&0.50&0.03&$-54.38$&0.75&0.00&1.00&0.50&0.02&\\
\cline{1-6}\cline{7-12}\cline{13-13}
$-38.17$&3.25&2.03&0.00&0.51&0.21&$-33.64$&3.26&2.05&0.00&0.51&0.22&$p_{3/2}$\\
$-36.60$&3.29&2.03&0.00&0.52&0.23&$-32.16$&3.38&2.03&0.00&0.52&0.26&\\
$-32.13$&3.70&2.07&0.00&1.48&1.00&$-27.66$&3.65&2.09&0.00&1.46&1.01&\\
$-31.12$&3.72&2.08&0.00&1.47&1.00&$-26.63$&3.75&2.09&0.00&1.48&1.00&\\
\cline{1-6}\cline{7-12}\cline{13-13}
$-27.61$&1.53&2.19&0.00&0.63&1.00&$-23.66$&1.39&2.03&0.00&0.63&0.98&$p_{1/2}$\\
$-26.26$&1.50&2.08&0.00&0.59&0.99&$-21.80$&1.48&2.09&0.00&0.61&1.00&\\
\cline{1-6}\cline{7-12}\cline{13-13}
$-18.37$&5.42&4.07&0.97&0.56&0.31&$-14.14$&5.81&4.25&0.97&0.60&0.34&$\alpha$ in $sd$-shell\\
$-17.25$&6.12&4.46&0.98&0.60&0.39&$-13.07$&5.90&4.30&0.97&0.62&0.39&\\
\cline{1-6}\cline{7-12}\cline{13-13}
$-11.26$&7.36&5.75&0.96&1.33&1.06&$-7.29$&7.32&5.77&0.98&1.38&1.05&$\approx[2113/2]$\\
\end{tabular}
\end{ruledtabular}
\end{table*}


The lower 20 orbits for 10 protons and 10 neutrons form the $^{20}\textrm{Ne}$ core 
and the last two orbits correspond to the valence proton and neutron around it.
In the $^{20}\textrm{Ne}$ core, the four nucleons in the $sd$-shell are not in the ideal $d_{5/2}$ orbits, 
but they form an $\alpha$ cluster at the surface of $^{16}\textrm{O}$. 
As a result, the intrinsic states of ${}^{22} \textrm{Na}$
is well deformed. The four nucleons in the $\alpha$ cluster do 
not contribute to the GT transitions 
because they form a spin-isospin saturated state. It is a different feature from the case of four nucleons
in the lowest Nilsson orbits in the $N=2$ shell in a deformed mean field.

On the other hand, the single-particle properties of the last two valence nucleons around the $^{20}$Ne core 
show nature of the spin-orbit favored Nilsson $[Nn_z\Lambda\Omega]=[2113/2]$ orbit in prolate deformation.
In the $T=0$ states, the $K=3$ band is the lowest because 
two $[2113/2]$ nucleons in the intrinsic spin $S=1$ state feel 
the attraction of the triplet-even nuclear interaction.
Thus, the intrinsic structure of 
the ground $K=3$ band of ${}^{22} \textrm{Na}$ is simply described by 
the $^{20}\textrm{Ne}$ core with two valence neutrons in the $[211+3/2]^p[211+3/2]^n$
configuration.

The deformation and single-particle properties of the intrinsic states of 
$K=0$ and $K=1$ bands are similar to those of the $K=3$ band. 
Also in the $K=0$ and $K=1$ bands, 
the $^{20}\textrm{Ne}$ core is formed by the lower 20 orbits for 10 protons and 10 neutrons.
The last two orbits for the valence proton and neutron around the core 
have the dominant $[211+3/2]$ component, but they also contain 
other minor components such as the $[211-3/2]$ and $[211-1/2]$ orbits. 
The $K=0$ and $K=1$ bands are produced from these minor components
by the $J^\pi$ and $K$ projections. Namely,
the $1_1^+0$ ($K=0$) state contains the $[211+3/2]^{p(n)}[211-3/2]^{n(p)}$
configuration of two nucleons coupling to $S=1$ with $S_z=0$.
On the other hand, the $1_2^+0$ ($K=1$) state is regarded as the $[211+3/2]^{p(n)}[211-1/2]^{n(p)}$
configuration of two nucleons with $S_z=1$.

With the similar analysis of the single-particle orbits for ${}^{22} \textrm{Ne}$, the intrinsic state of 
${}^{22} \textrm{Ne}(0_1^+1)$ is described by two $[2113/2]$-orbit neutrons 
in the $[211+3/2]^{n}[211-3/2]^{n}$ configuration with $K=0$ 
around the $^{20}$Ne core.


In the single-particle analysis, it is found that the 
structures of the low-lying states of $^{22}\textrm{Ne}$ and $^{22}\textrm{Na}$ 
are approximately described by Nilsson orbit configurations of two valence nucleons around the 
deformed $^{20}\textrm{Ne}$ core, which we call the ``$\Omega\Omega$-coupling scheme'' in this paper.
The GT transitions from 
${}^{22} \textrm{Ne}(0_1^+1)$ to $^{22}\textrm{Na}(1_{1,2}^+0)$ are mainly contributed 
by the transitions of two valence neutrons, $nn\to pn$, around the $^{20}\textrm{Ne}$ core.
In the $0_1^+1\to1_1^+0$ transition 
two valence neutrons 
$\Ket{n\uparrow n\downarrow}$ decay into $\Ket{p\uparrow n\downarrow}$ with $\Delta S_z=0$, 
whereas, in the $0_1^+1\to1_2^+0$ transition they decay into $\Ket{p\downarrow n\downarrow}$ with 
$\Delta S_z=\pm1$. In the $\Omega\Omega$-coupling scheme, the former $\Delta S_z=0$ and the latter 
$\Delta S_z=\pm1$ transitions correspond to the 
$[211+3/2]^{n}[211-3/2]^{n}\to [211+3/2]^{p(n)}[211-3/2]^{n(p)}$ and 
$[211+3/2]^{n}[211-3/2]^{n}\to [211+3/2]^{p(n)}[211-1/2]^{n(p)}$, respectively. 
Thus, the GT transition from ${}^{22} \textrm{Ne}(0_1^+1)$ is split into the spin non-flip and 
flip states in ${}^{22} \textrm{Na}$.

\subsection{Comparison of GT transitions of ${}^{10}\textrm{Be}\to{}^{10}\textrm{B}$ and ${}^{22}\textrm{Ne}\to{}^{22}\textrm{Na}$}
In order to give more general discussions of the low-energy GT transitions in deformed systems, 
we compared the GT transitions of ${}^{22}\textrm{Ne}\to{}^{22}\textrm{Na}$ with those of ${}^{10}\textrm{Be}\to{}^{10}\textrm{B}$
studied with the same method in the previous work \cite{Self2}, because 
${}^{10}\textrm{Be}$ and ${}^{10}\textrm{B}$ are also deformed nuclei in the $p$-shell with two valence nucleons
around the $2\alpha$ core.
In ${}^{10}\textrm{Be}\to{}^{10}\textrm{B}$, the strong GT transitions 
occur in valence two nucleons from a $nn$ pair to a $pn$ pair around the core.

\begin{figure}
\includegraphics[width=\hsize]{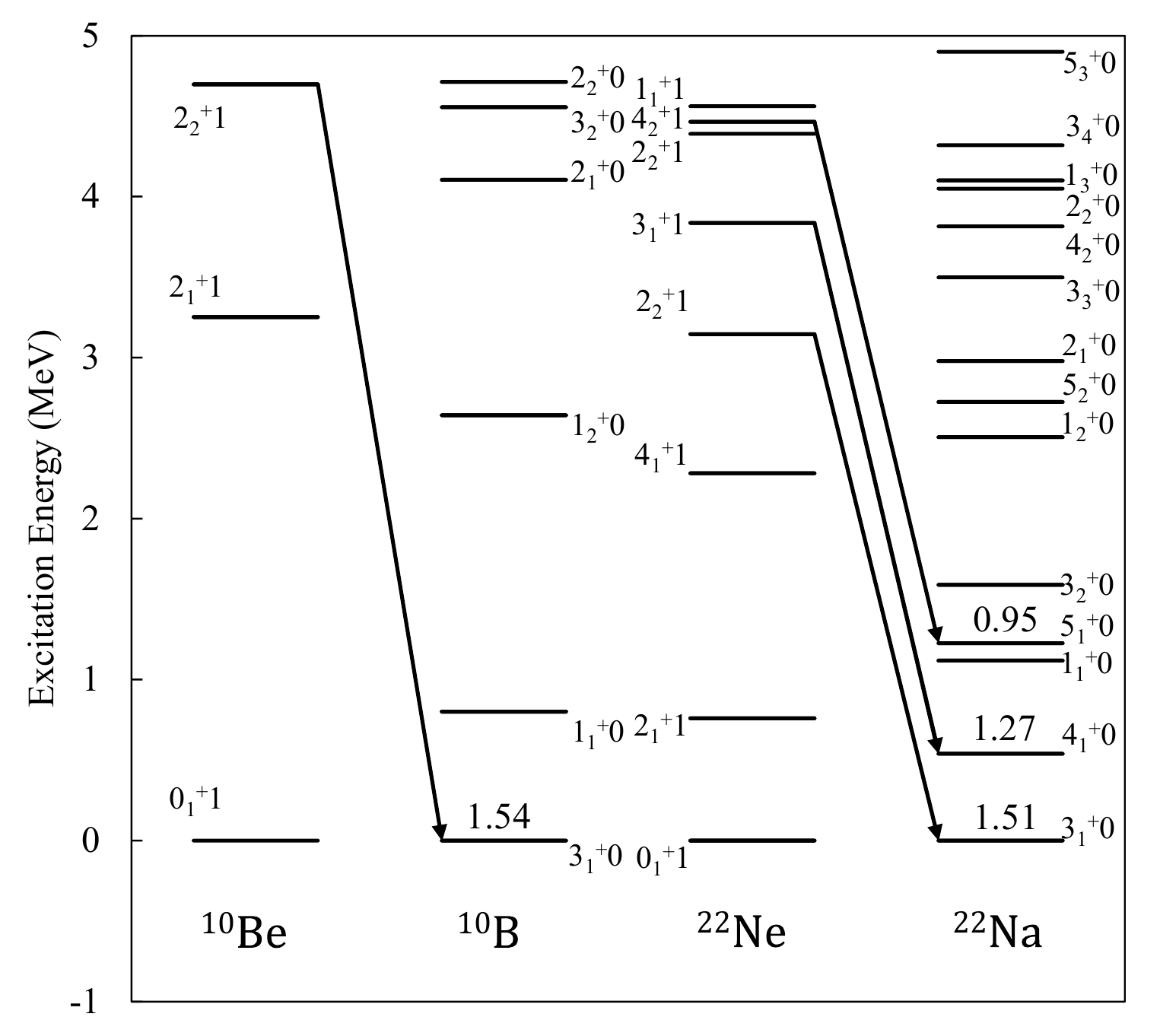}
\caption{The GT transitions $K=2\to K=3$ with large $B(\textrm{GT})$ values are shown with solid arrows. 
The energy is measured from each ground state.}
\label{GTSpecK3}
\end{figure}

\begin{figure}
\includegraphics[width=0.8\hsize]{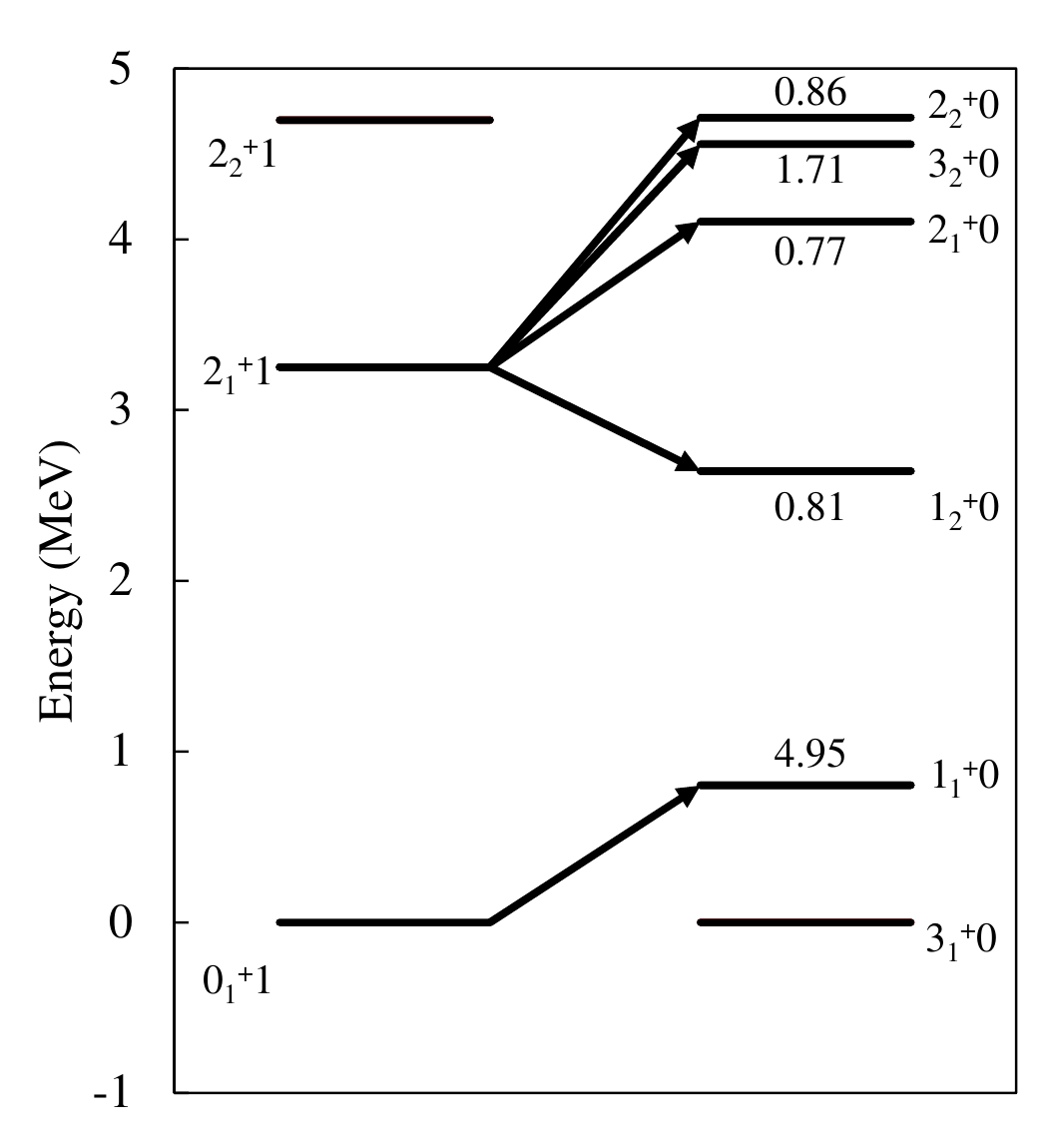}
\caption{The spectra of initial and final states in ${}^{10}\textrm{Be}\to{}^{10}\textrm{B}(T=0)$. 
The energy is measured from each ground state.
The states having large $B(\textrm{GT})$ are connected by arrows.}
\label{GTSpecB}
\end{figure}

In Figs.~\ref{GTSpecK3} and \ref{GTSpecB},
the energy spectra and $B(\textrm{GT})$ values in ${}^{10}\textrm{Be}$ and ${}^{10}\textrm{B}$ 
calculated with the $T\beta\gamma$-AMD+GCM are shown. 
The GT transitions are strong in ${}^{10}\textrm{Be}(0_1^+1)\to{}^{10}\textrm{B}(1_1^+0)$, ${}^{10}\textrm{Be}(2_1^+1)\to{}^{10}\textrm{B}(1_2^+0,2_{1,2}^+0,3_2^+0)$, 
and ${}^{10}\textrm{Be}(2_2^+1)\to{}^{10}\textrm{B}(3_1^+0)$.
The initial states of ${}^{10}\textrm{Be}$ are in the $K=0$ or $K=2$ band.
The final states in ${}^{10}\textrm{B}$ are 
regarded as the spin-isospin partner states of the $K=0$ or $K=2$ band members in ${}^{10}\textrm{Be}$
as is discussed in the previous paper \cite{Self2}.

The strong GT transition ${}^{10}\textrm{Be}(2_2^+1)\to {}^{10}\textrm{B}(3_1^+0)$ is regarded as the transition
from the $K=2$ side band to the ground $K=3$ band, which corresponds well to the GT transition 
of ${}^{22}\textrm{Ne}(2_2^+1)\to{}^{22}\textrm{Na}(3_1^+0)$.  
On the other hand, the GT transitions from the $K=0$ ground band of ${}^{10}\textrm{Be}$ show
different features from those of ${}^{22}\textrm{Ne}$. 
The GT transition from ${}^{10}\textrm{Be}(0_1^+1)$ is not split but concentrated on the single
${}^{10}\textrm{B}(1_1^+0)$ state because the final states in ${}^{10}\textrm{B}$ 
do not have definite $K$ quanta even though they have the deformed $2\alpha$ core with 
two valence nucleons. Instead, they have spatially developed deuteron-like $pn$ pairs weakly coupling 
with the $2\alpha$ core in the ``$LS$-coupling'' scheme rather than the $\Omega\Omega$-coupling scheme.

\begin{figure}
\includegraphics[width=\hsize]{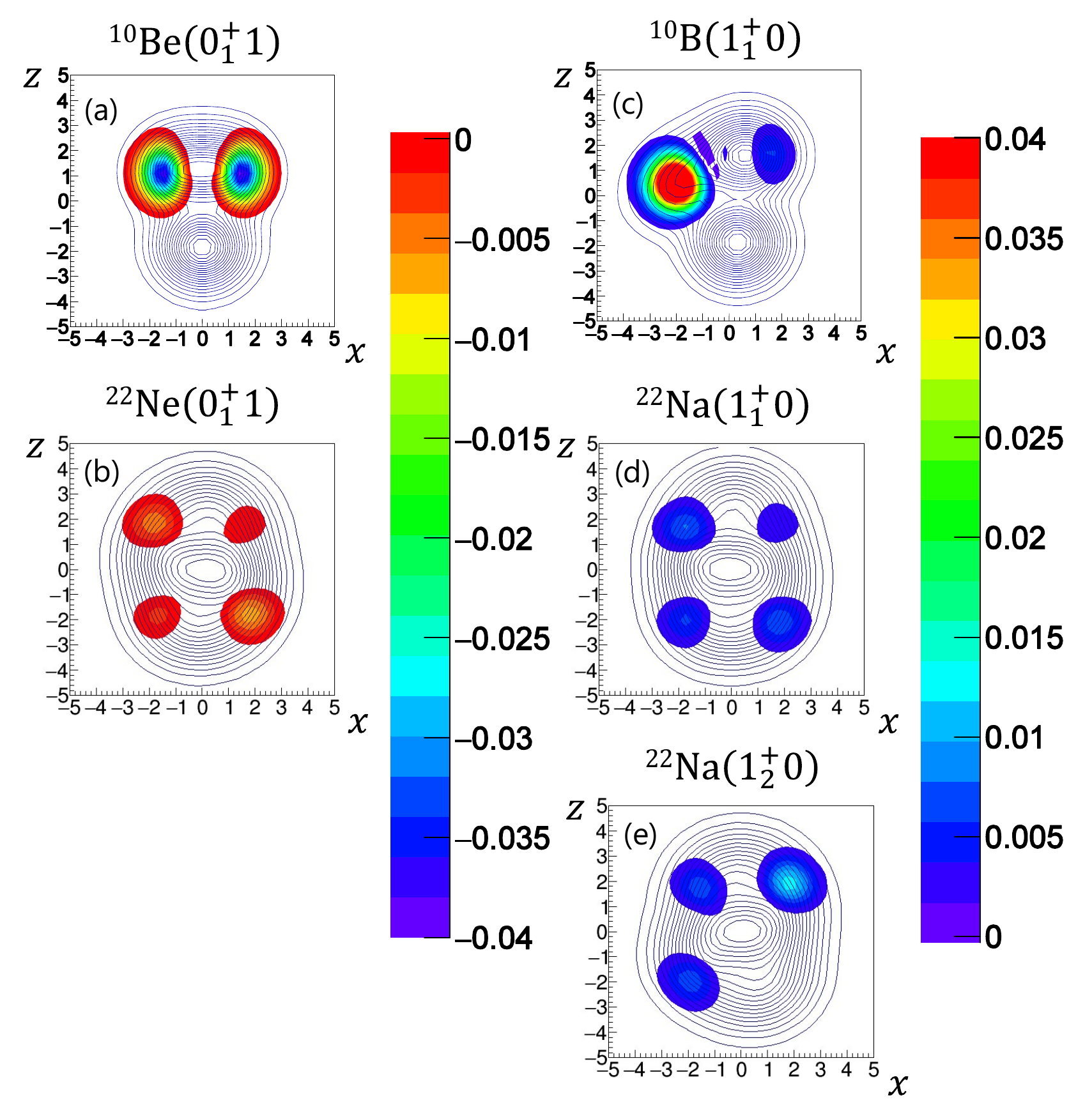}
\caption{The two-nucleon-pair density $\rho_{NN}({\mbox{\boldmath $r$}})$ of (a) ${}^{10} \textrm{Be}(0_1^+1)$, (b) ${}^{22} \textrm{Ne}(0_1^+1)$, (c) ${}^{10} \textrm{B}(1_1^+0)$, (d) ${}^{22} \textrm{Na}(1_1^+0)$, and (e) ${}^{22} \textrm{Na}(1_2^+0)$. The one-body density distribution $\rho({\mbox{\boldmath $r$}})$ is also shown by 
(blue) solid contour lines.
}
\label{density}
\end{figure}

In order to see spatial correlations of $NN$ pairs, we visualized the spatial distribution of the $S=1,T=0$ and $S=0,T=1$ $NN$ pairs 
with two-particle density ${\rho}_{ST}({\mbox{\boldmath $r$}})$ defined as 
\begin{eqnarray}
\rho_{ST}({\mbox{\boldmath $r$}}) = \frac{\Braket{\Phi^T(\beta,\gamma)|\hat{\rho}_{ST}({\mbox{\boldmath $r$}})|\Phi^T(\beta,\gamma)}}{\Braket{\Phi^T(\beta,\gamma)|\Phi^T(\beta,\gamma)}},\\
\hat{\rho}_{ST}({\mbox{\boldmath $r$}}) \equiv \sum_{ij}\hat{P}^S_{ij}\hat{P}^T_{ij}\delta({\mbox{\boldmath $r$}}-\hat{{\mbox{\boldmath $r$}}}_i)\delta({\mbox{\boldmath $r$}}-\hat{{\mbox{\boldmath $r$}}}_j),
\end{eqnarray}
where $\hat{P}^S_{ij}$ and $\hat{P}^T_{ij}$ are the spin and isospin projection operators for two nucleons \cite{Self2}. 
In Fig.~\ref{density}, we show $\rho_{NN}({\mbox{\boldmath $r$}})\equiv\rho_{10}({\mbox{\boldmath $r$}})-\rho_{01}({\mbox{\boldmath $r$}})$ for the major components of ${}^{10} \textrm{Be}$,${}^{10} \textrm{B}$,${}^{22} \textrm{Ne}$, and ${}^{22} \textrm{Na}$. Here, $\rho_{01}({\mbox{\boldmath $r$}})$ is subtracted to cancel contributions from the core nuclei.
In ${}^{10} \textrm{Be}$ and ${}^{10} \textrm{B}$, the $2\alpha$ cluster is elongated along the $z$-axis as seen in 
Figs.~\ref{density}(a) and (c). In the ${}^{10} \textrm{B}(1^+_1 0)$, 
the $T=0$ $pn$ pair distribution has a remarkable peak at $(x,z)=(-2,0)$ (fm) and shows the spatially developed  deuteron-like $pn$ pair
far from the $2\alpha$ core.
In contrast, the $nn$ and $pn$ pairs in ${}^{22} \textrm{Ne}$ and ${}^{22} \textrm{Na}$ are distributed at the surface of the deformed core and show no spatial development. 

\begin{figure}
\includegraphics[width=0.8\hsize]{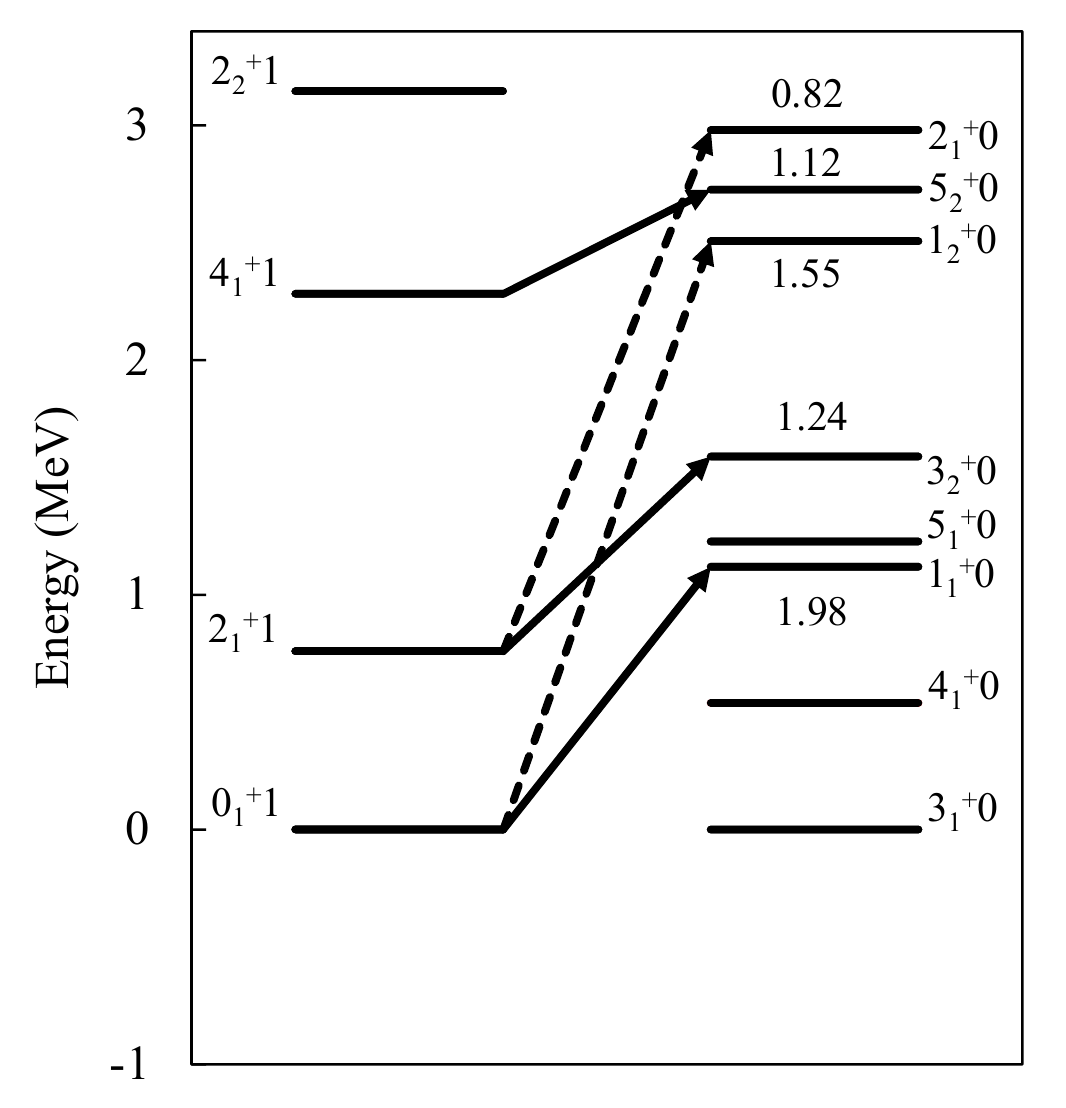}
\caption{The spectra of initial and final states in ${}^{22}\textrm{Ne}\to{}^{22}\textrm{Na}(T=0)$.
The energy is measured from each ground state. 
The states having large $B(\textrm{GT})$ are connected by arrows.
The solid arrows are $K=0\to K=0$ transitions and the dashed arrows are $K=0\to K=1$ ones.
}
\label{GTSpecNa}
\end{figure}

As a result of the $2\alpha$ cluster formation and the spatial development of the deuteron-like $pn$ pair, 
the $1_{1,2}^+0,2_{1,2}^+0,3_2^+0$ states of ${}^{10}\textrm{B}$ are constructed by the coupling
of the $S=1$ $pn$ pair with the orbital angular momentum $L$ of the $2\alpha$ core
as $[L=0,S=1]_{J=1}$ and $[L_\textrm{core}=2,S=1]_{J=1,2,3}$. 
Here, both $2_{1,2}^+0$ states contain $[L_\textrm{core}=2,S=1]_{J=2}$ component
because of the configuration mixing between the core rotation $L_\textrm{core}=2$ and $pn$ pair rotation $L_{pn}=2$ in the $J=2$ state \cite{Self2}.
The strong GT transition ${}^{10}\textrm{Be}(0_1^+1) \to {}^{10}\textrm{B}(1_1^+0)$
corresponds to $[L=0,S=0]_{J=0}\to[L=0,S=1]_{J=1}$, whereas 
the significant GT transitions of  ${}^{10}\textrm{Be}(2_1^+1)\to{}^{10}\textrm{B}(1_2^+0,2_{1,2}^+0,3_2^+0)$
are described by the transitions $[L_\textrm{core}=2,S=0]_{J=2}\to[L_\textrm{core}=2,S=1]_{J=1,2,3}$. 
In ${}^{10}\textrm{B}$, the intrinsic spin of the  $LS$-coupling $pn$ pair weakly couples with the core deformation. In such the case, the GT transition from the ground state ${}^{10}\textrm{Be}(0_1^+1)$ is not split but concentrated on the single state ${}^{10}\textrm{B}(1_1^+0)$ because both the spin flip and non-flip operators in the GT transition operator 
can contribute to the same final state. It is a consequence of the $SU(4)$-symmetry 
of the $LS$-coupling $pn$ pair in ${}^{10}\textrm{B}$. 

Differently from the GT transition of ${}^{10}\textrm{Be}(0_1^+1)\to {}^{10}\textrm{B}(1_1^+0)$, 
the GT splitting occurs in ${}^{22}\textrm{Ne}(0_1^+1)\to{}^{22}\textrm{Na}(1_{1,2}^+0)$ 
(see Fig.~\ref{GTSpecNa}).
As already discussed previously, 
the origin of the splitting is that the final states of ${}^{22}\textrm{Na}$ have specific $K$ quanta 
because of the 
$\Omega\Omega$-coupling $pn$ pair around the $^{20}$Ne core. 
The GT transitions to the final $K=1$ and $K=0$ bands 
occur 
in the $nn\to pn$ decays with $\Delta S_z=0$ and $\Delta S_z=\pm 1$  in the intrinsic frames, respectively.
The key point is that ${}^{22}\textrm{Na}(1^+_{1,2}0)$ have the 
$\Omega\Omega$-coupling $pn$ pair with the $SU(4)$-symmetry breaking, and 
${}^{10}\textrm{B}(1^+_10)$ has the  $LS$-coupling 
$pn$ pair with the $SU(4)$-symmetry.

In order to discuss the roles of $pn$ correlation and core deformation in the GT splitting (or fragmentation) phenomena, 
we performed a further analysis of the GT transitions for the artificially prepared final states with 
the $pn$ pairs around deformed cores in the $LS$-coupling limit and in the $\Omega\Omega$-coupling case, 
and in the $jj$-coupling limit.
To this end, we changed the strength $u_{ls}$ of the spin-orbit interaction $V_{LS}$
as $u_{ls}=\lambda u^\textrm{default}_{ls}$ with the enhancement factor $\lambda$ from the default strength
$u^\textrm{default}_{ls}=1300$~MeV, and performed
the GCM calculation of ${}^{10}\textrm{B}$ and ${}^{22}\textrm{Na}$. 
In the GCM calculation, we used the bases $\{|\Phi^{\pi T}(\beta_i,\gamma_i)\rangle\}_i$ obtained with the default spin-orbit strength.
The $u_{ls}\to 0$ limit corresponds to the $LS$-coupling scheme with the $SU(4)$-symmetry, whereas 
in the large $u_{ls}$ limit the system goes to spherical states with $jj$-coupling nucleons.
In the intermediate case of  $u_{ls}$, $\Omega\Omega$-coupling $NN$ pairs appear around the deformed core. 
By controlling the enhancement factor in the range of $\lambda=0-2$, we 
discuss how the GT transitions are fragmented in the change from $LS$-coupling regime to the $jj$-coupling regime.

\begin{figure}
\includegraphics[width=0.5\hsize]{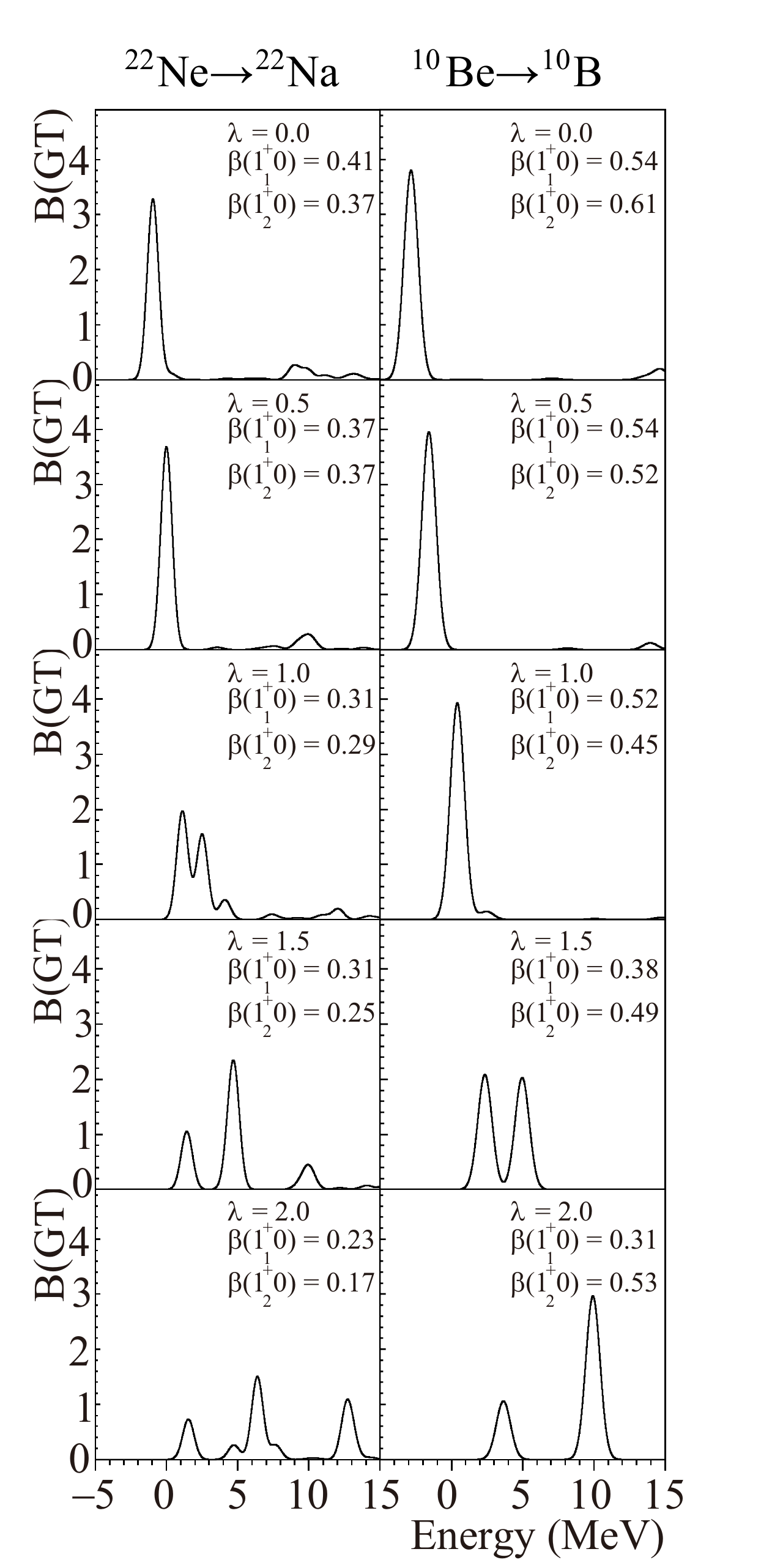}
\caption{
The $B(\textrm{GT})$ spectra obtained by the calculations with 
the modified spin-orbit strengths with $\lambda=0.0, 0.5, 1.0, 1.5, 2.0$. 
The $\lambda=1.0$ corresponds to the default strength. 
Each spectrum is smeared by gaussian with $\sigma=0.4$ so as to normalize the peak hight to the $B(\textrm{GT})$ value
for the case of an isolate peak.
For each  $\lambda$, 
the energies are measured from ${}^{22} \textrm{Na}(3_1^+0)$ and  ${}^{10} \textrm{B}(3_1^+0)$, respectively.
The left and right panels show $B(\textrm{GT};{}^{22} \textrm{Ne}\to{}^{22} \textrm{Na})$ and $B(\textrm{GT};{}^{10} \textrm{Be}\to{}^{10} \textrm{B})$, respectively.
}
\label{LSdep}
\end{figure}

In Fig.~\ref{LSdep}, we show the $B(\textrm{GT};{}^{22}\textrm{Ne}(0_1^+1)\to{}^{22}\textrm{Na}(1_n^+0))$ spectra for $\lambda=0.0, 0.5, 1.0, 1.5$ and $2.0$
in comparison with the $B(\textrm{GT};{}^{10}\textrm{Be}(0_1^+1)\to{}^{10}\textrm{B}(1_n^+0))$ spectra. 
Let us first discuss the GT transitions of ${}^{22}\textrm{Ne}(0_1^+1)\to{}^{22}\textrm{Na}(1_n^+0)$.
In the $\lambda=0.0$ case for the $LS$-coupling limit, the $pn$ pair has the $SU(4)$-symmetry and the GT strength is concentrated on the 
single lowest state with a large fraction 54.8\% of the GT sum rule value 6. 
As $\lambda$ increases, the GT strengths from ${}^{22}\textrm{Ne}(0_1^+1)$ are split into a few 
$1^+0$ states in the low-energy region.
At the default strength $\lambda=1.0$, the $LS$-coupling $pn$ pair 
around the $^{20}\textrm{Ne}$ changes to  
the $\Omega\Omega$-coupling pair, and the GT strengths are split mainly to two states ${}^{22}\textrm{Na}(1_{1,2}^+0)$.  
Significant GT strengths still exist in the low-energy region: The sum of the strengths for these two states 
exhausts 58.8\% of the sum rule value. This corresponds to partial breaking of the $SU(4)$-symmetry in $sd$-shell nucleons
because this symmetry is broken only in the $pn$ pairs but not in the $\alpha$ cluster. 
With further increase of $\lambda$ to $\lambda=2.0$,  
the deformation parameter becomes small as $\beta=0.31\to 0.23$ and 
the GT strengths are fragmented into many $1^+0$ states.
The major peak position rises up to the higher energy and the strength function is widely distributed.
In the large $\lambda$ case, the $\alpha$ cluster is broken by the strong spin-orbit force and 
the system goes to the $jj$-coupling regime, in which six nucleons in the $sd$-shell contribute to the GT transitions.

Next we look into the $B(\textrm{GT};{}^{10}\textrm{Be}(0_1^+1)\to{}^{10}\textrm{B}(1_n^+0))$ spectra and
compare them with 
${}^{22}\textrm{Ne}(0_1^+1)\to{}^{22}\textrm{Na}(1_n^+0)$.
Also in the ${}^{10}\textrm{Be}(0_1^+1)\to{}^{10}\textrm{B}(1_n^+0)$, 
we find similar behavior of the GT splitting with the $SU(4)$-symmetry breaking.
In the small $\lambda$ case for the $LS$-coupling limit,
the GT strength from the ${}^{10}\textrm{Be}(0_1^+1)$ is concentrated on the ${}^{10}\textrm{B}(1_1^+0)$ with $B(\textrm{GT})\approx5.0$ which almost exhausts the sum rule value because of 
the $SU(4)$-symmetry of the $NN$ pair around the $2\alpha$ core.   
As $\lambda$ increases, the GT peak is split into two states $1_{1,2}^+0$ and shifted toward the high-energy region.
One of the remarkable differences 
from ${}^{22}\textrm{Ne}(0_1^+1)\to{}^{22}\textrm{Na}(1_n^+0)$ is that,
in the case of ${}^{10}\textrm{Be}(0_1^+1)\to{}^{10}\textrm{B}(1_n^+0)$,  the 
splitting occurs not at $\lambda=1.0$ (the default spin-orbit strength) but at 
$\lambda=1.5$
because the $NN$ pairs around the $2\alpha$ core favor the $LS$-coupling scheme.
It means that, in the realistic system at $\lambda=1.0$, the $SU(4)$-symmetry in the $pn$ pair still remains and 
the GT transition is concentrated on the single low-lying $^{10}\textrm{B}(1_1^+0)$. 

In the present analysis, we found an universal feature of the GT fragmentation phenomena in deformed systems. 
There are two types of the fragmentation mechanism of the GT strengths. 
One is the GT splitting in the $K=0,1$ bands because of the $SU(4)$-symmetry breaking in the $pn$ pairs 
around largely deformed core with the spin-isospin saturated configurations.
This corresponds to the partial breaking of the $SU(4)$-symmetry.
The other is the GT fragmentation in $jj$-coupling shell orbits 
in the weakly deformed system. The ${}^{22}\textrm{Ne}(0_1^+1)\to{}^{22}\textrm{Na}(1_{1,2}^+0)$
is the former case of the partial breaking phase, whereas 
the ${}^{10}\textrm{Be}(0_1^+1)\to{}^{10}\textrm{B}(1_{1}^+0)$ is close to the ideal 
$SU(4)$-symmetry phase with no GT splitting.

\section{Summary and outlook}
\label{summary}
We have investigated the Gamow-Teller transitions of ${}^{22}\textrm{Ne}\to{}^{22}\textrm{Na}$ with the $T\beta\gamma$-AMD+GCM
in order to discuss the relation between strong GT transitions and $pn$ pair formation in the prolately deformed $N=Z$ odd-odd nuclei.
The splitting of the GT strengths from ${}^{22}\textrm{Ne}(0_1^+1)\to{}^{22}\textrm{Na}(1_{1,2}^+0)$ is found
reproducing the experimental data in the mirror transitions; ${}^{22}\textrm{Mg}(0_1^+1)\to{}^{22}\textrm{Na}(1_{1,2}^+0)$.
This GT splitting is understood by introducing the ``$\Omega\Omega$-coupling scheme'' of the valence $NN$ pair around the spin-isospin saturated $^{20}$Ne core.
By analyzing the major components of the $K$ band heads, we have found the final states ($1_1^+0,3_2^+0,5_2^+0$) are in the $K=0$ band and other states ($1_2^+0,2_1^+0$) are in the $K=1$ band.
The single-particle orbits for the valence two particles of the $1_1^+0$ ($K=0$) state show that this $K$ quantum is produced by $[211+3/2]^{p(n)}[211-3/2]^{n(p)}$
configuration with the intrinsic spins coupled to $S=1$ with $S_z=0$.
On the other hand, those of the $1_2^+0$ ($K=1$) state have the major $[211+3/2]^{p(n)}[211-1/2]^{n(p)}$ configuration with $S_z=\pm 1$.
This fact shows the GT splitting is caused by the $SU(4)$-symmetry breaking of the $pn$ pair into the $\Omega\Omega$-coupling scheme 
producing $K$ quanta in the intrinsic frame.
For the final states in the $K=0$ and $K=1$ bands with the
$S_z=0$ and $S_z=\pm 1$ $pn$ pairs, the GT transitions 
occur in the $nn\to pn$ decays with $\Delta S_z=0$ (spin-nonfilp) and $S_z=\pm 1$ (spin-filp) in the intrinsic frame, respectively. 

We have also compared the GT transitions in ${}^{22}\textrm{Ne}\to{}^{22}\textrm{Na}$ with those in ${}^{10}\textrm{Be}\to{}^{10}\textrm{B}$.
The GT splitting in ${}^{22}\textrm{Ne}(0_1^+1)\to{}^{22}\textrm{Na}(1_{1,2}^+0)$ occurs as a result of the $SU(4)$-symmetry breaking in the $pn$ pair 
around largely deformed core with the spin-isospin saturated configurations.
On the other hand, the ${}^{10}\textrm{Be}(0_1^+1)\to{}^{10}\textrm{B}(1_{1}^+0)$ is close to the 
$SU(4)$-symmetry phase with no GT splitting
because the $pn$ pair in ${}^{10}\textrm{B}$ is spatially developed and contains both the $K=0,1$ quanta though the $2\alpha$ core is also deformed.

\section*{Acknowledgments}
The computational calculations of this work were performed using the supercomputers in the Yukawa Institute for theoretical physics, Kyoto University. 
This work was supported by JSPS KAKENHI Grant Numbers 16J05659, 26400270, and 18K03617.

\end{document}